\documentstyle[galley,epsf,epsfig]{mn2e}

\title[Accurate Fundamental Parameters or A, F, and G-type Supergiants in the Solar Neighbourhood]
   {Accurate Fundamental Parameters for A, F, and G-type Supergiants in the Solar Neighbourhood}

\author[L.S. Lyubimkov et al.]
       {Leonid~S.~Lyubimkov,$^1$\thanks{E-mail: lyub@crao.crimea.ua (LSL); dll@astro.as.utexas.edu (DLL)}
       David~L.~Lambert,$^2$$^\star$
     Sergey~I.~Rostopchin,$^2$ \and Tamara~M.~Rachkovskaya$^1$ and Dmitry~B.~Poklad$^1$\\
       $^1$Crimean Astrophysical Observatory, Ukraine\\
       $^2$The W.J. McDonald Observatory, The University of Texas at Austin, USA\\}
\date{Accepted Received ; in original form  }
 
\pagerange{\pageref{firstpage}--\pageref{lastpage}} \pubyear{}

\begin{document}

\label{firstpage}

\maketitle

\begin{abstract}
The following  parameters are determined for 63 Galactic supergiants in 
the solar neighbourhood: effective temperature $T_{\rm eff}$, 
surface gravity $\log g$,  iron abundance $\log\epsilon$(Fe), 
microturbulent parameter $V_t$, mass $M/M_\odot$, age $t$ and distance $d$. 
A significant improvement in the accuracy of the determination of $\log g$ and,  
all parameters dependent on it, is obtained through application of van Leeuwen's
(2007)
rereduction of the {\it Hipparcos} parallaxes. The typical error in the $\log g$ 
values  is now $\pm$0.06 dex for supergiants with distances $d < 300$ pc 
and $\pm$0.12 dex for supergiants with $d$ between 300 and 700 pc;  the mean error 
in $T_{\rm eff}$ for these stars is $\pm$120 K. For supergiants with $d > 700$ pc 
parallaxes are uncertain or unmeasurable, so typical errors in their $\log g$ 
values are 0.2--0.3 dex. 

A new $T_{\rm eff}$ scale for A5--G5 stars of luminosity classes Ib--II is
presented. Spectral subtypes and luminosity classes of several stars are
corrected. Combining  the $T_{\rm eff}$ and $\log g$ with evolutionary
tracks, stellar masses and ages are determined; a majority of the sample has
masses between 4$M_\odot$ and 15$M_\odot$ and, hence, their
progenitors were early to middle B-type main sequence stars.

Using Fe\,{\sc ii} lines, which are insensitive to departures from LTE, 
the microturbulent parameter $V_t$ and the iron abundance $\log\epsilon$(Fe)
are determined from high-resolution spectra. 
The parameter $V_t$ is correlated with gravity: $V_t$ increases with
decreasing $\log g$.
The mean iron abundance for the 48 supergiants with distances $d$ $<$ 700 pc
is $\log\epsilon$(Fe)=7.48$\pm$0.09, a value close to the solar
value of 7.45$\pm$0.05, and thus
the local supergiants and the Sun have the same metallicity.
\end{abstract}


\section{Introduction}

In this new series of papers, we shall obtain and comment upon
chemical compositions of a large sample of A-,F-, and G-type
supergiants. An initial step in an abundance analysis is the
determination of the principal stellar fundamental parameters:
the effective temperature $T_{\rm eff}$, the surface gravity
$\log g$, the iron abundance $\log\epsilon$(Fe), and the microturbulent
parameter $V_t$. In now traditional fashion, a model atmosphere is
computed for a given set of fundamental parameters and applied to
the analysis of the observed stellar spectrum. Iteration is
essentially necessary because the derived composition may differ from that
assumed in the construction of the model atmosphere. Additionally, 
analysis of the spectrum may provide new information on the
fundamental parameters. 

In this paper, the sample of supergiants is
introduced and their fundamental parameters derived. For many
stars, this is not the first presentation of estimates of their
fundamental parameters. A new discussion appeared vital because
inspection of the literature   shows generally divergent
results; large uncertainties about $T_{\rm eff}$ and $\log g$
(particularly) translate to inaccurate estimates of the
elemental abundances. As one illustrative example, we show in Table 1 
published $T_{\rm eff}$ and $\log g$ for $\alpha$ Per (HR 1017)
where the ranges are  340 K and 1.4 dex.  Lyubimkov et al. (2009)
note that the scatter for the F0 Ib star $\alpha$ Lep (HR 1865)
is 500 K in $T_{\rm eff}$ and 1.2 dex in $\log g$. An even more
extreme spread of 2400 K in $T_{\rm eff}$ is reported by Schiller \&
Przybilla (2008) for the A2 Ia supergiant $\alpha$ Cyg (Deneb, HR 7924):
the spread in $\log g$ from their collated references was 0.5 dex.
A salutary lesson to be drawn from these examples of bright well-studied
supergiants is that fundamental parameters of supergiants in the
literature deserve a careful redetermination if one is seeking
accurate abundances.

\begin{table}
\centering
 \begin{minipage}{70mm}
  \caption{The $T_{\rm eff}$ and $\log g$ values derived by various authors for the bright supergiant $\alpha$ Per (F5 Ib)}
  \begin{tabular}{lll}
  \hline
   $T_{\rm eff}$, K & $\log g$ & Authors\\
 \hline
 6250 & 1.8 & Parsons (1967) \\
 6250 & 0.90 & Luck \& Lambert (1985) \\
 6500 & 1.5 & Klochkova \& Panchuk (1988)\\
 6250 & 1.20 & Gonzalez \& Lambert (1996)\\
 6270 & -- & Evans et al. (1996)\\
 6200 & 0.60 & Andreievsky et al. (2002)\\
 6541 & 2.0 & Kovtyukh et al. (2008)\\
 6350 & 1.90 & present work \\
 \hline
  \end{tabular}
 \end{minipage}
\end{table}

Here, we consider not only familiar
spectroscopic and photometric methods for determining the
fundamental parameters but also a determination of the
gravity $\log g$ that is based on the stellar parallax. The
{\it Hipparcos} parallaxes from the original 
catalogue (ESA 1997) have been significantly improved by van Leeuwen (2007).
In particular, the parallaxes for many bright
supergiants  have their errors reduced by 4--5 times, a reduction
that allows the stellar parallax to be a competitive method for
estimating $\log g$.

Our initial sample of 67 stars was
reduced to 63 after $T_{\rm eff}$ and $\log g$ estimates had been
obtained and four stars shown to be
misclassified  in {\it The Bright Star Catalogue}
(Hoffleit \& Warren 1991). For these 63 supergiants, we determined
the parameters --
$T_{\rm eff}$, $\log g$, $V_t$, and $\log\epsilon$(Fe) -- as well as the 
stellar mass $M/M_\odot$, age $t$, and distance $d$.

Determination of these fundamental parameters is a prerequisite for
the abundance analyses pursued in this series of papers to
examine questions primarily of stellar
evolution (say, mixing arising from rotation and deep convection)
and secondarily of Galactic chemical evolution.
Several abundance anomalies in atmospheric
composition of supergiants have been found and widely
attributed to mixing between the interior and the atmosphere.
For example, the carbon abundance shows a
tendency to be underabundant 
(e.g., Luck \& Lambert 1985; Venn 1995a,b; Venn \& Przybilla
2003). Nitrogen, as might be expected for C-depleted stars, is
overabundant. Sodium appears to be overabundant (Boyarchuk \&
Lyubimkov 1983; Lyubimkov 1994; Andrievsky et al. 2002). The lithium
abundance varies greatly from star-to-star (Luck 1977). These anomalies
with respect to a star's initial composition are presumed traceable to
the mixing into the atmosphere of nuclear-processed material from
the stellar interior. There are two possible episodes of mixing:
(i) rotationally-induced mixing in the rapidly-rotating main sequence
(MS) B-type progenitor of the supergiant; and (ii) the deep convective
mixing, the so-called first dredge-up, that occurs towards the end of
the first crossing the Hertzsrpung gap when the star is a K- or M-type
supergiant and before a return to earlier spectral types.
Since rotationally-induced mixing in the MS progenitors should
be confined to the most rapidly-rotating stars and effects of the first 
dredge-up seen only in stars that have completed a first-crossing of
the Hertzsrpung gap, one expects the strength of the abundance
anomalies (Li, C, N, Na etc.)  to vary from absent to
marked  across a large sample of
A, F, and G supergiants. If this expectation is not confirmed, i.e.,
an anomaly is present in all stars, the conclusion must be that
episodes (i) and (ii) are incompletely understood or another
process exists for affecting the surface compositions.

\section{Stellar selection and observations}

Stars were selected from {\it The Bright Star Catalogue} (Hoffleit \& Warren 1991). 
The sample of 67 stars comprised stars classified in the catalogue as
having spectral types from A5 to G8 and luminosity classes
I and II (i.e., supergiants and bright giants); three stars were
subsequently shown to be luminosity class IV or V objects. Cepheid
variables were not included in the sample. Stars inaccessible from
the McDonald Observatory were  not considered.

Binaries were excluded too, as far as possible. Nevertheless, some 
programme supergiants may have faint companions. The bright 
F-type binary supergiant $\pi$~Sgr (HR 7264), which consists of two components 
of equal brightness was included. This star was interesting for 
us, because its parameters were determined earlier by various authors; in 
particular, $\pi$ Sgr is one of three programme stars, for which the effective 
temperature $T_{\rm eff}$ has been inferred from the infrared flux method (see below).

High-resolution spectra were obtained of the 67 stars using the
cross-dispersed echelle coud\'{e} spectrograph at the Harlan J.
Smith 2.7-m telescope of the W.J. McDonald Observatory (Tull et al.
1995). Essentially complete spectral coverage of the optical
spectrum was obtained at high S/N ratio. Reductions of the
CCD  spectral images were performed using standard IRAF routines.
 
Measurements on the reduced spectra included the equivalent
widths of the Balmer lines H$\beta$ and H$\gamma$ and many
Fe\,{\sc ii} lines. 
Measured equivalent widths
$W$(H$\beta$) and $W$(H$\gamma$) were corrected for the weak extended
wings (Lyubimkov et al. 2000). 
The Balmer lines are too blended to be measured
in the cool G supergiants. When measuring Fe\,{\sc ii} lines, we described
their profiles by Gaussian.

\section{Determination of Effective Temperature  and Surface Gravity}

Photometric indices and the Balmer lines constitute a major
component of 
the determination of the parameters $T_{\rm eff}$ and $\log g$.
The approach adapted from that applied earlier by us to
B-type stars (Lyubimkov et al. 2002). 
was discussed in detail
by Lyubimkov et al. (2009) who applied it to four supergiants from the
present sample. 
The two reddening-independent indices are the $Q=(U-B)-0.72(B-V)$
index from the
$UBV$ Johnson photometric system and the $[c_1]=c_1-0.20(b-y)$
from the   $uvby$
Str\"{o}mgren system. The Balmer lines are introduced into the
determination through the $\beta$-index and the equivalent widths of
H$\beta$ and H$\gamma$.  

Each of these quantities provides a locus in the ($T_{\rm eff},\log g$)
plane. The locus is set by the observed
value of an index and its calibration from model atmospheres and
synthetic spectra.  
Observed indices $Q$, $[c_1]$ and $\beta$ are taken from Hauck \&
Mermilliod's (1998) catalogue. 
Computations of $W$(H$\beta$) and $W$(H$\gamma$) are taken from
Kurucz (1993). Predicted indices $Q$ and $[c_1]$ are taken from
Castelli \& Kurucz (2003) and predicted $\beta$ values from
Castelli \& Kurucz (2006). These predictions are based on
on ATLAS model atmospheres. 

The infrared flux method (IRFM) has proven to be a powerful
and accurate way to determine $T_{\rm eff}$. Blackwell \&
Lynas-Gray (1998) give results for two of our supergiants
($\alpha$ Aqr and $\beta$ Aqr).  Ram\'{\i}rez \& Mel\'{e}ndez
(2005) present the $T_{\rm eff}$ for $\pi$ Sgr.

The trigonometrical parallax $\pi$ offers a way to supplement the
loci from photometry and Balmer lines. The well known expression
is
\begin {eqnarray}
\log d = -5.25 + 0.5\log M/M_\odot +2\log T_{\rm eff} -0.5\log g \nonumber \\
       +0.2\,m_V -0.2\,A_V +0.2\,BC,
\end{eqnarray}
where $d$ is the distance in parsec, $M/M_\odot$ is the star's mass in
solar masses, $m_V$ is the visual magnitude, $A_V$ is the interstellar
extinction, and $BC$ is the bolometric correction. In the application
here, we rewrite the expression
\begin{eqnarray}
\log g-\log M/M_\odot -0.4\,BC = -10.50 + 4\log T_{\rm eff} \nonumber \\
                             +2\log\pi +0.4\,m_V -0.4\,A_V.
\end{eqnarray}
 
For a given value of $T_{\rm eff}$, the right-hand side of this
equation has a fixed value. Solving the equation for $\log g$ 
requires the stellar mass and the BC which are estimated from stellar
evolutionary tracks (Claret 2004) and predictions of the BC from model
atmospheres. By repeating this procedure for different $T_{\rm eff}$
the ($T_{\rm eff},\log g$) loci is traced out. This valuable
application of  stellar parallax is discussed further by
Lyubimkov et al. (2009) who note that the method is not seriously
compromised by the presence of loops in evolutionary tracks back
across the Hertzsprung gap for stars with masses greater than about 
6 $M_\odot$. 

To illustrate our combined use of photometric indices, the Balmer
lines and the stellar parallax, we show the ($T_{\rm eff},\log g$) loci
for $\pi$ Sgr (F2 II) in Figure 1, and  $\alpha$ Aqr (G2 Ib) and
$\beta$ Aqr (G0 Ib) in Figure 2. Lyubimkov et al. (2009) show the  
similar figure for $\alpha$ Lep (F0 Ib). 
To the just mentioned loci, we add the $T_{\rm eff}$ from the IRFM.
One notes the excellent convergence of the various loci in the case of
$\pi$ Sgr. The locus from the parallax $\pi$ is especially valuable in
determining the gravity. We adopt $T_{\rm eff} = 6590\pm50$ K and
$\log g$ = $2.21\pm0.05$.   

\begin{figure}
\epsfxsize=8truecm
\epsffile{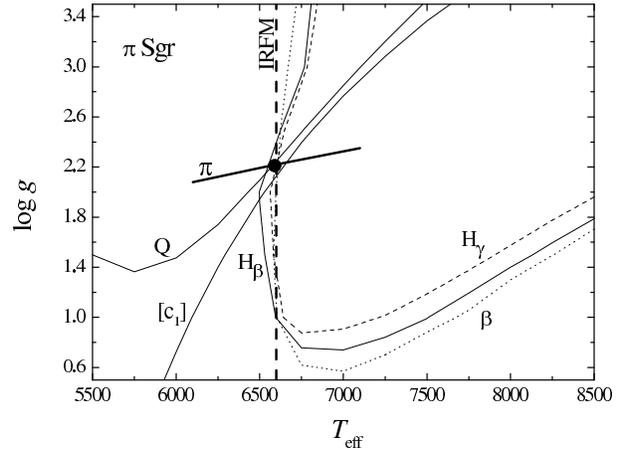}
\caption{The ($T_{\rm eff},\log g$) plane for $\pi$ Sgr (F2 II) showing the
loci provided from the indices $Q$, $[c_1]$, $\beta$, and the equivalent
widths $W$(H$\beta$) and $W$(H$\gamma$), and a short segment of the
locus derived from the parallax $\pi$.  The vertical dashed line corresponds to
the effective temperature from the IRFM. The filled circle presents the adopted parameters.}
\end{figure}
\begin{figure}
\epsfxsize=8truecm
\epsffile{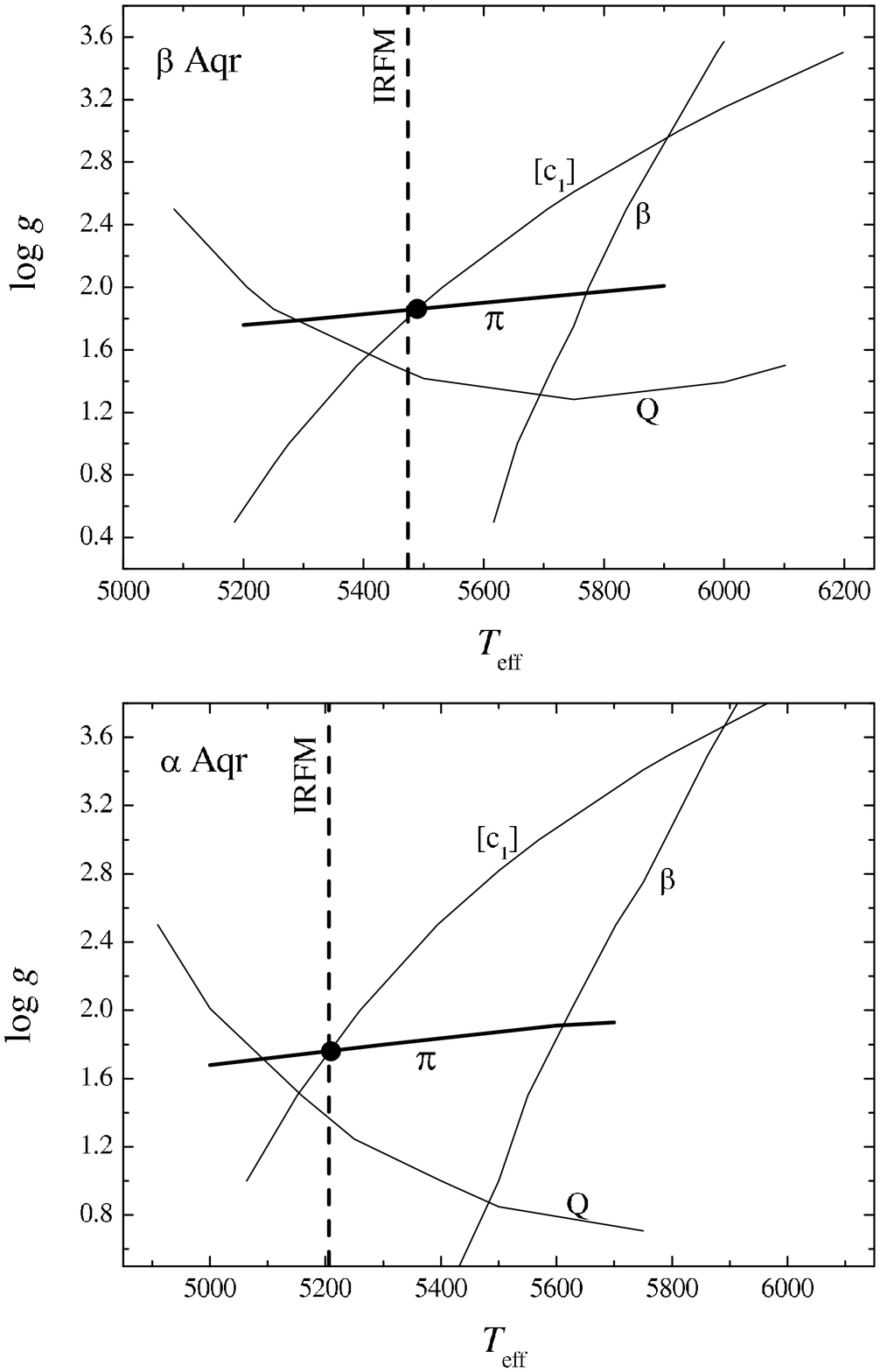}
\caption{The ($T_{\rm eff},\log g$) plane for $\beta$ Aqr (G0 Ib) (top panel)
 and $\alpha$ Aqr (G2 Ib) (bottom panel) showing the
loci provided from the indices $Q$, $[c_1]$, $\beta$, 
 and a short segment of the
locus derived from the parallax $\pi$.  The vertical dashed line corresponds to
the effective temperature from the IRFM. 
The filled circle shows the adopted parameters.}
\end{figure}

It should be noted once again that the star $\pi$~Sgr is a binary system, 
which consists of two components of equal brightness with the separation 
of 0.093$"$ between them (see {\it The Bright Star Catalogue}). Nevertheless, 
Fig.1 shows no traces of duplicity; note, in particular, an excellent 
agreement with the IRFM. The parameters $T_{\rm eff}$ and $\log g$ and the chemical 
composition of the star $\pi$~Sgr as a binary system have been analyzed by 
Lyubimkov and Samedov (1987). They found that the ignoring of the star 
duplicity can lead to some underestimation of element abundances. 

For the cooler G-type stars $\alpha$~Aqr and
$\beta$~Aqr, the IRFM $T_{\rm eff}$  effectively crosses the $[c_1]$ and 
$\pi$ loci at their intersection. 
 Guided by this result, we shall
derive $T_{\rm eff}$ and $\log g$ for G supergiants from the 
combination of the  $[c_1]$
index and the parallax $\pi$ for those stars lacking a $T_{\rm eff}$ 
measurement from the IRFM. 
Addition of the $Q$ index to the mix with equal weight given to
the several loci would suggest a slightly lower $T_{\rm eff}$ and $\log g$.
The  $\beta$ index
for these G-type supergiants  provides loci
that do not intersect the area of convergence provided by the $Q$,
$[c_1]$ loci and the $T_{\rm eff}$ from the IRFM. 
\begin{table*}
 \centering
 \begin{minipage}{140mm}
  \caption{Parameters of relatively near supergiants ($d < 700$ pc) }
  \begin{tabular}{ccccccccc}
 \hline
 HR & HD & Name & SP        & $m_v$, & $\pi$, & $d$, pc & $T_{\rm eff}$, K & $\log g$\\
    &    &      & (BS Cat.) & mag    & mas \\
\hline
27   & 571    & 22 And       & (F2 II)   & 5.03 & 2.63  $\pm$ 0.28 & 380   $\pm$ 40  & 6270 $\pm$ 150 & 2.10 $\pm$ 0.08\\
157  & 3421   &              & G2.5 IIa  & 5.44 & 3.86  $\pm$ 0.26 & 259   $\pm$ 17  & 5130 $\pm$ 150 & 2.15 $\pm$ 0.07\\
292  & 6130   &              & F0 II     & 5.96 & 1.63  $\pm$ 0.53 & 613   $\pm$ 199 & 6880 $\pm$ 100 & 2.05 $\pm$ 0.21\\
461  & 9900   &              & (G5 II)   & 5.55 & 1.89  $\pm$ 0.41 & 529   $\pm$ 115 & 4430 $\pm$ 100 & 1.18 $\pm$ 0.14\\
792  & 16780  &              & G5 II     & 6.31 & 2.52  $\pm$ 0.55 & 397   $\pm$ 87  & 5020 $\pm$ 100 & 2.09 $\pm$ 0.16\\
849  & 17818  &              & G5 Iab:   & 6.25 & 1.86  $\pm$ 0.48 & 538   $\pm$ 139 & 5020 $\pm$ 100 & 1.73 $\pm$ 0.18\\
1017 & 20902  & $\alpha$ Per & F5 Ib     & 1.82 & 6.43  $\pm$ 0.17 & 156   $\pm$ 4   & 6350 $\pm$ 100 & 1.90 $\pm$ 0.04\\
1135 & 23230  & $v$ Per      & F5 II     & 3.78 & 5.87  $\pm$ 0.22 & 170   $\pm$ 6   & 6560 $\pm$ 50  & 2.44 $\pm$ 0.03\\
1242 & 25291  &              & F0 II     & 5.04 & 1.59  $\pm$ 0.27 & 629   $\pm$ 107 & 6815 $\pm$ 100 & 1.87 $\pm$ 0.11\\
1270 & 25877  &              & G8 IIa    & 6.30 & 2.34  $\pm$ 0.44 & 427   $\pm$ 80  & 5060 $\pm$ 150 & 1.91 $\pm$ 0.13\\
1303 & 26630  & $\mu$ Per    & G0 Ib     & 4.18 & 3.63  $\pm$ 0.20 & 275   $\pm$ 15  & 5380 $\pm$ 100 & 1.73 $\pm$ 0.06\\
1327 & 27022  &              & (G5 IIb)  & 5.27 & 10.21 $\pm$ 0.37 & 98    $\pm$ 4   & 5440 $\pm$ 200 & 2.89 $\pm$ 0.07\\
1603 & 31910  & $\beta$ Cam  & G1 Ib-IIa & 4.03 & 3.77  $\pm$ 0.21 & 265   $\pm$ 15  & 5300 $\pm$ 100 & 1.79 $\pm$ 0.06\\
1740 & 34578  & 19 AUR       & A5 II     & 5.05 & 1.57  $\pm$ 0.33 & 637   $\pm$ 134 & 8300 $\pm$ 100 & 2.10 $\pm$ 0.25\\
1829 & 36079  & $\beta$ Lep  & (G5 II)   & 2.84 & 20.34 $\pm$ 0.18 & 49.2  $\pm$ 0.4 & 5450 $\pm$ 100 & 2.60 $\pm$ 0.03\\
1865 & 36673  & $\alpha$ Lep & F0 Ib     & 2.60 & 1.47  $\pm$ 0.15 & 680   $\pm$ 70  & 6850 $\pm$ 80  & 1.34 $\pm$ 0.07\\
2000 & 38713  &              & (G2 Ib-II)& 6.17 & 4.46  $\pm$ 0.45 & 224   $\pm$ 23  & 5000 $\pm$ 250 & 2.45 $\pm$ 0.14\\
2453 & 47731  & 25 Gem       & G5 Ib     & 6.44 & 1.58  $\pm$ 0.53 & 633   $\pm$ 212 & 4900 $\pm$ 100 & 1.70 $\pm$ 0.23\\
2693 & 54605  & $\delta$ CMa & F8 Ia     & 1.84 & 2.02  $\pm$ 0.38 & 495   $\pm$ 93  & 5850 $\pm$ 150 & 1.00 $\pm$ 0.14\\
2786 & 57146  &              & G2 Ib     & 5.30 & 2.59  $\pm$ 0.29 & 386   $\pm$ 43  & 5260 $\pm$ 150 & 1.90 $\pm$ 0.09\\
2833 & 58526  &              & G3 Ib     & 5.98 & 2.67  $\pm$ 0.40 & 375   $\pm$ 56  & 5380 $\pm$ 150 & 2.21 $\pm$ 0.12\\
2881 & 59890  &              & G3 Ib     & 4.60 & 2.17  $\pm$ 0.24 & 461   $\pm$ 51  & 5300 $\pm$ 100 & 1.66 $\pm$ 0.08\\
3045 & 63700  & $\zeta$ Pup  & G6 Iab-Ib & 3.34 & 2.70  $\pm$ 0.21 & 370   $\pm$ 29  & 4880 $\pm$ 150 & 1.21 $\pm$ 0.09\\
3073 & 64238  & 10 Pup       & F1 Ia     & 5.70 & 2.96  $\pm$ 0.37 & 338   $\pm$ 42  & 6670 $\pm$ 50  & 2.61 $\pm$ 0.08\\
3102 & 65228  & 11 Pup       & F7 II     & 4.20 & 6.23  $\pm$ 0.23 & 161   $\pm$ 6   & 5690 $\pm$ 200 & 2.17 $\pm$ 0.08\\
3183 & 67456  &              & A 5 II    & 5.35 & 2.08  $\pm$ 0.32 & 481   $\pm$ 74  & 8530 $\pm$ 100 & 2.67 $\pm$ 0.10\\
3188 & 67594  & $\xi$ Mon    & G2 Ib     & 4.36 & 3.08  $\pm$ 0.27 & 325   $\pm$ 28  & 5210 $\pm$ 100 & 1.75 $\pm$ 0.07\\
3229 & 68752  & 20 Pup       & G5 II     & 5.00 & 3.75  $\pm$ 0.30 & 267   $\pm$ 21  & 5130 $\pm$ 100 & 2.04 $\pm$ 0.06\\
3459 & 74395  &              & G1 Ib     & 4.64 & 4.23  $\pm$ 0.27 & 236   $\pm$ 15  & 5370 $\pm$ 100 & 2.08 $\pm$ 0.06\\
4166 & 92125  & 37 LMi       & (G2.5 IIa)& 4.69 & 5.64  $\pm$ 0.25 & 177   $\pm$ 8   & 5475 $\pm$ 50  & 2.36 $\pm$ 0.04\\
4786 & 109379 & $\beta$ Crv  & G5 II     & 2.65 & 22.41 $\pm$ 0.19 & 44.6  $\pm$ 0.4 & 5100 $\pm$ 80  & 2.52 $\pm$ 0.03\\
5143 & 119035 &              & G5 II:    & 5.21 & 6.04  $\pm$ 0.44 & 166   $\pm$ 12  & 5190 $\pm$ 200 & 2.75 $\pm$ 0.09\\
5165 & 118605 & 83 Vir       & G0 Ib-IIa & 5.57 & 3.95  $\pm$ 0.28 & 253   $\pm$ 18  & 5430 $\pm$ 100 & 2.37 $\pm$ 0.06\\
6081 & 147084 & o Sco        & A5 II     & 4.55 & 3.72  $\pm$ 0.54 & 269   $\pm$ 39  & 8370 $\pm$ 200 & 2.12 $\pm$ 0.15\\
6536 & 159181 & $\beta$ Dra  & G2 Ib-IIa & 2.79 & 8.58  $\pm$ 0.10 & 116.6 $\pm$ 1.4 & 5160 $\pm$ 150 & 1.86 $\pm$ 0.04\\
6978 & 171635 & 45 Dra       & F7 Ib     & 4.78 & 1.54  $\pm$ 0.17 & 649   $\pm$ 72  & 6000 $\pm$ 50  & 1.70 $\pm$ 0.07\\
7164 & 176123 &              & G3 II     & 6.39 & 2.99  $\pm$ 0.53 & 334   $\pm$ 59  & 5200 $\pm$ 200 & 2.25 $\pm$ 0.15\\
7264 & 178524 & $\pi$ Sgr    & F2 II     & 2.89 & 6.41  $\pm$ 0.43 & 156   $\pm$ 10  & 6590 $\pm$ 50  & 2.21 $\pm$ 0.05\\
7456 & 185018 &              & G0 Ib     & 5.99 & 2.70  $\pm$ 0.42 & 370   $\pm$ 58  & 5550 $\pm$ 150 & 2.06 $\pm$ 0.11\\
7542 & 187203 &              & F8 Ib-II  & 6.47 & 2.66  $\pm$ 0.48 & 376   $\pm$ 68  & 5750 $\pm$ 150 & 2.15 $\pm$ 0.15\\
7795 & 194069 &              & (G5 III+A)& 6.40 & 2.48  $\pm$ 0.34 & 403   $\pm$ 55  & 4870 $\pm$ 150 & 2.00 $\pm$ 0.08\\
7796 & 194093 & $\gamma$ Cyg & F8 Ib     & 2.24 & 1.78  $\pm$ 0.27 & 562   $\pm$ 85  & 5790 $\pm$ 100 & 1.02 $\pm$ 0.10\\
7834 & 195295 & 41 Cyg       & F5 II     & 4.02 & 4.15  $\pm$ 0.17 & 235   $\pm$ 9   & 6570 $\pm$ 80  & 2.32 $\pm$ 0.08\\
8232 & 204867 & $\beta$ Aqr  & G0 Ib     & 2.91 & 6.07  $\pm$ 0.23 & 165   $\pm$ 6   & 5490 $\pm$ 100 & 1.86 $\pm$ 0.05\\
8313 & 206859 & 9 Peg        & G5 Ib     & 4.34 & 3.53  $\pm$ 0.22 & 283   $\pm$ 18  & 4910 $\pm$ 100 & 1.58 $\pm$ 0.06\\
8412 & 209693 &              & G5 Ia     & 6.38 & 3.52  $\pm$ 0.43 & 284   $\pm$ 35  & 5280 $\pm$ 150 & 2.35 $\pm$ 0.09\\
8414 & 209750 & $\alpha$ Aqr & G2 Ib     & 2.95 & 6.20  $\pm$ 0.19 & 161   $\pm$ 5   & 5210 $\pm$ 100 & 1.76 $\pm$ 0.04\\
8692 & 216206 &              & G4 Ib     & 6.24 & 2.42  $\pm$ 0.38 & 413   $\pm$ 65  & 4960 $\pm$ 100 & 1.90 $\pm$ 0.11\\
\end{tabular}
\end{minipage}
\end{table*}

For most of the programme G supergiants, we found $T_{\rm eff}$ to be higher from $\beta$ than from $[c_1]$. 
One may suppose that the cause can be connected with 
some systematic errors in the computed or observed $\beta$ values. As mentioned above, we 
used Castelli \& Kurucz's (2006) computed $\beta$ values based on ATLAS model atmospheres. 
Recently \"Onehag et al. (2009) calculated colour indices in the $uvby$-$\beta$ photometric system 
for stars with $T_{\rm eff} = 4500–-7000$ K, using a new generation of MARCS model atmospheres. 
(The code MARCS was constructed specially for calculations of models for relatively cool stars, 
see Gustafsson et al. 2008). We compared their and Castelli \& Kurucz's $\beta$ values for $\log g = 2.0$, 
the lowest surface gravity in \"Onehag et al.'s computations. The difference between two sets of data 
seems to be negligible in the case of F stars with $T_{\rm eff} > 6000$ K and less than 1 per cent in the 
case of G stars with $T_{\rm eff} > 4500$ K. Such a small difference cannot change markedly the temperatures 
$T_{\rm eff}$ derived from the $\beta$-index. 

\begin{table*}
 \centering
 \begin{minipage}{140mm}
  \caption{Parameters of distant supergiants ($d > 700$ pc) }
  \begin{tabular}{cccccclcc}
 \hline
 HR & HD & SP        & $m_V$, & $\pi$, & $d$, pc & $A_V$ & $T_{\rm eff}$, K & $\log g$\\
    &    & (BS Cat.) & mag    & mas \\
\hline
 207 & 4362  & G0 Ib    & 6.42 & 1.07  $\pm$ 0.50 & 935  $\pm$ 440 & 0.56 (0.68) & 5220 $\pm$ 100 & 1.55  $\pm$  0.31\\
 825 & 17378 & A5 Ia    & 6.28 & -0.85 $\pm$ 0.48 & 2700 $^{a)}$   & 2.62        & 8570 $\pm$ 160 & 1.18  $\pm$  0.13\\
2597 & 51330 & F2 Ib-II & 6.28 & 1.07  $\pm$ 0.53 & 935  $\pm$ 460 & 0.46        & 6710 $\pm$ 100 & 2.02  $\pm$  0.34\\
2839 & 58585 & A8 I-II  & 6.05 & 0.75  $\pm$ 0.48 & 1330 $\pm$ 850 & 0.30        & 7240 $\pm$ 150 & 1.92  $\pm$  0.25\\
2874 & 59612 & A5 Ib    & 4.85 & 1.06  $\pm$ 0.28 & 940  $\pm$ 250 & 0.37        & 8620 $\pm$ 200 & 1.78  $\pm$  0.21\\
2933 & 61227 & F0 II    & 638  & 1.27  $\pm$ 0.62 & 790  $\pm$ 380 & 0.93 (0.89) & 6690 $\pm$ 150 & 2.02  $\pm$  0.33\\
3291 & 70761 & F3 Ib    & 5.90 & -0.27 $\pm$ 0.30 & 2900 $^{a)}$   & 0.25        & 6600 $\pm$ 100 & 1.25  $\pm$  0.30\\
6144 & 148743 & A7 Ib   & 6.48 & 0.75  $\pm$ 0.43 & 1330 $\pm$ 760 & 0.66        & 7400 $\pm$ 400 & 1.80  $\pm$  0.40\\
7014 & 172594 & F2 Ib   & 6.45 & 1.00  $\pm$ 0.50 & 1000 $\pm$ 500 & 1.84        & 6760 $\pm$ 100 & 1.66  $\pm$  0.34\\
7094 & 174464 & F2 Ib   & 5.84 & 1.17  $\pm$ 0.35 & 855  $\pm$ 260 & 1.19        & 6730 $\pm$ 200 & 1.75  $\pm$  0.21\\
7387 & 182835 & F3 Ib   & 4.69 & 1.14  $\pm$ 0.27 & 880  $\pm$ 210 & 1.14 (0.84) & 6700 $\pm$ 120 & 1.43  $\pm$  0.15\\
7770 & 193370 & F5 Ib   & 5.18 & 1.04  $\pm$ 0.21 & 960  $\pm$ 190 & 0.52        & 6180 $\pm$ 100 & 1.53  $\pm$  0.13\\
7823 & 194951 & F1 II   & 6.41 & 0.99  $\pm$ 0.41 & 1010 $\pm$ 420 & 0.82        & 6760 $\pm$ 100 & 1.92  $\pm$  0.27\\
7847 & 195593 & F5 Iab  & 6.20 & 0.96  $\pm$ 0.38 & 1040 $\pm$ 410 & 1.80        & 6290 $\pm$ 100 & 1.44  $\pm$  0.26\\
7876 & 196379 & A9 II   & 6.20 & 0.17  $\pm$ 0.25 & 1740 $^{a)}$   & 0.61        & 7020 $\pm$ 100 & 1.66  $\pm$  0.10\\
\hline\\
\end{tabular}
\\$^a) $\ These distances $d$ are evaluated from the $T_{\rm eff}$ and $\log g$ values (see text).\\
\end{minipage}
\end{table*}

As far as possible errors in the observed $\beta$ values are concerned, we estimated that, for instance, 
in the case of the stars $\beta$ Aqr and $\alpha$ Aqr (Figure 2) their $\beta$ values (2.621 and 2.606, respectively) 
should be decreased by about 0.02 in order to lead $T_{\rm eff}$ in agreement with temperatures inferred from 
$[c_1]$. One may see from Hauck \& Mermilliod's (1998) catalogue that, in fact, there is a scatter 
between measurements of various authors; in particular, Arellano Ferro et al. (1990) provide the 
appreciably lower value $\beta = 2.585$ for both above-mentioned supergiants. However, a possibility of 
{\it systematic} underestimates in the observed $\beta$ values for G supergiants remains unclear.

\section{Effective Temperatures and Surface Gravities}

\subsection{Supergiants with $d < 700$ pc}

Inspection of parallaxes of the programme stars in van Leeuwen's (2007)
catalogue showed that the parallax spans a wide range from 25 mas down to 
unmeasurably small values (mas =  milliarcsecond). 
There is a natural boundary near by $\pi$ = 1.5 mas providing a  separation 
between relatively near supergiants with the reliable $\log g$ values and 
distant supergiants with the less reliable $\log g$ values. 
We consider the first more numerous group  that consists of 48 stars with
distances $d <700$ pc. 

In Table 2, we give the HR and HD numbers of these stars, their 
spectral classification according to the BS catalogue, 
visual magnitude $m_V$, parallax $\pi$ and distance $d$ 
obtained from relation $d = 1/\pi$. 
Next,  we present in Table 2 the derived parameters 
$T_{\rm eff}$ and $\log g$, as well as the errors of their determination. 
Note that spectral classifications given in brackets  are unreliable;  
revised  spectral types are given below. 

The accuracy of van Leeuwen's parallaxes for stars listed in Table 2 
is rather high, so we may expect a corresponding high accuracy  for 
our $\log g$ values. Table 2 confirms this expectation:  
the average error for the  $\log g$ values for all 48 stars is $\pm$0.10 dex. 
If the selection is further limited to stars with $d < 300$ pc,
we find the mean error $\pm$0.06 dex in $\log g$ (23 objects), while for the 
stars with $d$ between 300 and 700 pc the error is $\pm$0.12 dex (25 objects). 
The average error in $T_{\rm eff}$ for all the stars is about $\pm$120 K.

\subsection{Supergiants with $d > 700$ pc}

Table 3 provides  findings for supergiants with $d > 700$ pc, 15 stars in all. 
Their parallaxes are small ($\pi < $1.3 mas) and are often comparable 
with errors in $\pi$.  Moreover, the reported $\pi$ are negative 
for two stars (HR 825 and 3291),  i.e.,  unmeasurable. 
Therefore, for HR 825 and 3291, as well as for the star HR 7876 with a
very small parallax, we could not use the parallax in construction of the 
($T_{\rm eff},\log g$)  diagrams. As a result, the error in $\log g$ 
for half the stars listed in Table 3 is large, namely about $\pm$0.3 dex 
attaining $\pm$0.4 dex for HR 6144. 
It should be noted that for the three above-mentioned supergiants, HR 825, 3291 
and 7876, where the parallax  is unmeasurable,  we evaluated their distance $d$ 
from equation (1) basing on the derived $T_{\rm eff}$ and $\log g$ values.

\begin{figure*}
\epsfxsize=8truecm
\epsfig{file=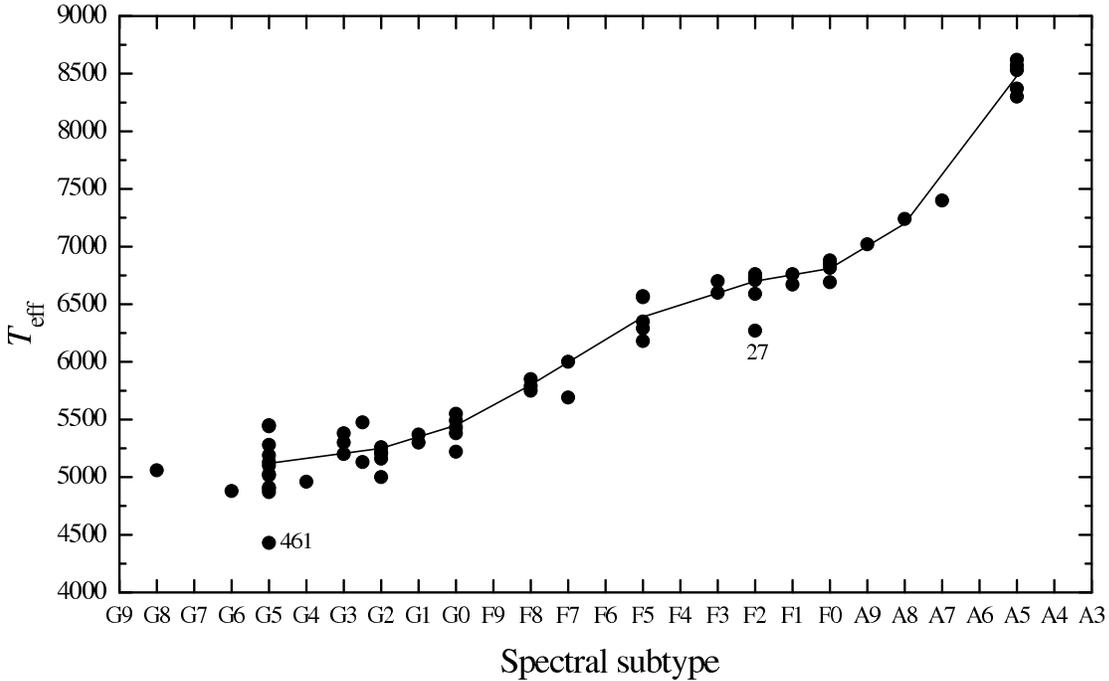,width=9cm,angle=270}
\caption{The $T_{\rm eff}$ scale for AFG stars of luminosity classes Ib--II.
Outliers HR~27, 461, and 6144 were omitted in constructing the mean
relation (solid line). }
\end{figure*}

Equation (2) includes the interstellar extinction $A_V$, so a question arises: 
how can uncertainties in $A_V$ affect the $\log g$ determination from parallaxes? 
This question is of special interest for distant supergiants, so we show in 
Table 3 our $A_V$ values for these stars. Moreover, we provide as well the $A_V$ values 
for three common stars (in brackets) from Kovtyukh et al. (2008), inferred from 
theirs $E(B-V)$ through relation $\ A_V = 3.12~E(B-V)$. One sees that there is good 
agreement for two stars, but for the third star, HR~7387, a marked difference, 
0.30, takes place. We redetermined $\log g$ for HR~7387 from equation (2) adopting 
Kovtyukh et al.'s value $\ A_V = 0.84$. As a result, $\log g$ increases by 0.08 dex that 
is significantly less than an error in the derived $\log g$ value. It should be noted 
that the difference of 0.30 is rather great; for most of common stars (22 in all) 
Kovtyukh et al.'s $A_V$ values differ from ours less than by 0.20. We concluded that 
uncertainties in $A_V$ affect slightly the surface gravities $\log g$ derived from parallaxes. 
(Comparison of our $T_{\rm eff}$, $\log g$ and $V_t$ values with Kovtyukh et al.'s parameters see below).

\subsection{Spectral Classification and Fundamental Parameters}

Tables 2 and 3 contain accurate $T_{\rm eff}$s for 63 supergiants;
four of the original sample of 67 were excluded (see below).
These temperatures allow us to construct a $T_{\rm eff}$ -- Spectral
type calibration for supergiants and luminous giants. The
calibration is shown in Figure 3 where the line corresponds to the
scale given in Table 4.

  \begin{table}
 \centering
 \begin{minipage}{70mm}
  \caption{Effective temperatures for supergiants and luminous giants }
  \begin{tabular}{cccc}
 \hline
  Spectral type & $T_{\rm eff}$ & Spectral type & $T_{\rm eff}$\\
                &      (K)  &               &      (K) \\
\hline
A5 & 8480 & F8 & 5800\\
A8 & 7200 & G0 & 5450\\
F0 & 6810 & G2 & 5250 \\
F2 & 6700 & G5 & 5120 \\
F5 & 6390 & & \\
\hline
\end{tabular}
\end{minipage}
\end{table}

The surface gravity is, as expected,  different for stars of
luminosity classes Ib and II. From Tables 2 and 3 we find that
late A-type and F-type supergiants of class Ib have a mean
$\log g$ = 1.6 and a typical range of 1.2--2.0. Class II A- and F-type
stars have $\log g$  generally between 2.0 and 2.4. In the case of
G-type stars, the typical $\log g$ are 1.5--2.0 for class Ib
and 2.0--2.6 for class II. Thus, the value $\log g$ =2.0 seems to be
the boundary between classes Ib and II.

\subsection{Incorrect Spectral Classifications}

Four of our 67 stars seem to have an erroneous spectral
classification in the BS catalogue; they are not AFG supergiants.
 These stars are listed in Table 5.
\begin{table*}
 \centering
 \begin{minipage}{150mm}
  \caption{Stars with erroneous classification in the BS Catalogue (excluded from further analysis) }
  \begin{tabular}{ccccccccccc}
 \hline
 HR & HD  & $m_V$, & $\pi$, & $d$, pc & $T_{\rm eff}$, K & $\log g$ & Sp & Sp & $M/M_\odot$ \\
    &     & mag    &  mas   &         &                  &          & BS Cat. & corrected & \\
\hline
1746 & 34658  & 5.34 & 16.14 $\pm$ 0.39 & 62.0  $\pm$ 1.5 & 6570 $\pm$ 80  & 3.58 $\pm$ 0.03 & F5 II & F5 IV-V & 2.0\\
2768 & 56731  & 6.32 & 7.56  $\pm$ 0.44 & 132   $\pm$ 8   & 7570 $\pm$ 50  & 3.58 $\pm$ 0.04 & A9 II & A9 IV-V & 2.3\\
8718 & 216756 & 5.91 & 25.66 $\pm$ 0.34 & 39.0  $\pm$ 0.5 & 6600 $\pm$ 50  & 4.07 $\pm$ 0.01 & F5 II & F5 V    & 1.5\\
2636 & 52611  & 6.20 & 5.68  $\pm$ 0.56 & 176   $\pm$ 17  & 4120 $\pm$ 100 & 1.78 $\pm$ 0.03 & G5 II & K2 Ib   & 1.5\\
\hline
\end{tabular}
\end{minipage}
\end{table*}

 They are among the fainter objects in our sample ($m_V = 5.3 - 6.3$ mag)
and yet are close to the Sun with distances from 39 to 176 pc. Three
of the stars, namely HR 1746, 2768, and 8718, are in the BS catalogue
as A- or F-type class II stars. However, luminous giants at such close
distances cannot be so faint. Therefore, their status as class II
stars is immediately suspect. The surface gravities in Table 3 definitely
show that these stars are dwarfs or subgiants. Table 5 provides
both the catalogue and revised classifications. We exclude these stars from
further discussion. 
The fourth star in Table 5, HR 2636, appears too cool to be a G5 II
star although the gravity confirms the luminosity class. We suggest it is
a K2 Ib star. 

Judged by their $T_{\rm eff}$ and $\log g$, several other stars appear to
require some adjustment to their spectral classification as given in the
BS Catalogue. For example, HR 27 and 461 are obvious outliers in
Figure 4. We list in Table 6 stars for which we suggest a revised
classification.

\subsection{Comparisons with the literature}

Of the published compilations of fundamental parameters for
A-, F-, and G-type Ib and II stars, we select for comparison
that by Kovtyukh et al. (2008) because these authors report a
high accuracy for their $T_{\rm eff}$ and $\log g$. Figure 4 presents
the comparisons for 22 F and G stars in common. There is fine agreement
between our and their $T_{\rm eff}$ for the G supergiants 
(see Figure 4 upper panel for $T_{\rm eff} < 5800$ K)  
but in the case of F stars Kovtyukh et al.'s  temperatures are
systematically hotter. The discrepancy is 650--740 K for the
three F0 supergiants (HR 1242, 1865, 2933).

\begin{figure}
\epsfxsize=8truecm
\epsffile{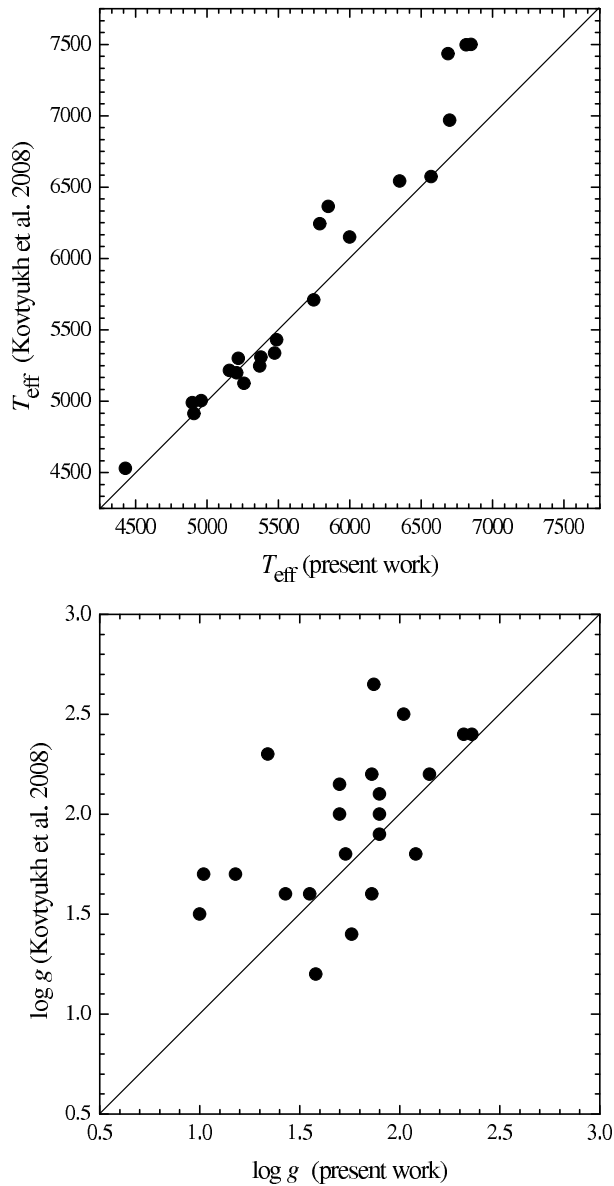}
\caption{Comparisons of our parameters $T_{\rm eff}$ and $\log g$ with
data of Kovtyukh et al. (2008). The solid stright line in each panel  denotes
perfect agreement between the sets of data. }
\end{figure}

Kovtyukh et al. determined $T_{\rm eff}$ by the method suggested by
Kovtyukh (2007) involving ratios of the central depths of specially
selected pairs of spectral lines of different excitation potential.
The method has two important features. First, it was calibrated using
$T_{\rm eff}$ values from the literature for a number of supergiants.
Thus, it does not provide an independent method of determining $T_{\rm eff}$
and can contain systematic errors carried over from the calibrating
stars. Second, the method does not always provide a unique estimate.
For example, in the case of $\alpha$ Lep (HR~1865) mentioned above, the result for
$T_{\rm eff}$ depends on the initial guess for the temperature
range, i.e. the choices $T_{\rm eff} > 7000$ K and $T_{\rm eff} < 7000$ K
lead to different conclusions, that is $T_{\rm eff} = 7510$ K and
$T_{\rm eff} = 6950$ K, respectively (Lyubimkov et al. 2009). 
Kovtyukh et al.'s published value was $T_{\rm eff} = 7500$ K whereas
our value is $T_{\rm eff} = 6850\pm80$ K (Table 2), a value only   
100 K cooler than the lower value from Kovtyukh et al.'s method.

\begin{table}
 \centering
 \begin{minipage}{85mm}
  \caption{Correction of the BS Catalogue classification for some programme stars }
  \begin{tabular}{ccccc}
 \hline
 HR & $T_{\rm eff}$, K & $\log g$ & Sp & Sp  \\
    &                  &          & BS Catal. & corrected  \\
\hline
27   & 6270 $\pm$ 150 & 2.10 $\pm$ 0.08 & F2 II    & F5 II\\
461  & 4430 $\pm$ 100 & 1.18 $\pm$ 0.14 & G5 II    & K0 Ia\\
1327 & 5440 $\pm$ 200 & 2.89 $\pm$ 0.07 & G5 II    & G0 II\\
1829 & 5450 $\pm$ 100 & 2.60 $\pm$ 0.03 & G5 II    & G0 II\\
2000 & 5000 $\pm$ 250 & 2.45 $\pm$ 0.14 & G2 Ib-II & G5 II\\
4166 & 5475 $\pm$ 50  & 2.36 $\pm$ 0.04 & G2.5 IIa & G0 II\\
7795 & 4870 $\pm$ 150 & 2.00 $\pm$ 0.08 & G5 III+A & G9 Ib-II\\
\hline
\end{tabular}
\end{minipage}
\end{table}
Surface gravities were estimated by Kovtyukh et al. from the
condition of ionization equilibrium using lines of Fe\,{\sc i} and
Fe\,{\sc ii}. Such $\log g$ values are dependent on the adopted
$T_{\rm eff}$. Comparison of $\log g$ values is of interest
because our results are dependent on the application of the
trigonmetrical parallax. The comparison with Kovtyuth et al.
is made in the lower panel of Figure 4. There is a scatter of
about $\pm$0.3 dex between the data sets. Not suprisingly,
the difference in the $\log g$ values is much larger
(0.5--0.9 dex) for the three F0  supergiants for which our
$T_{\rm eff}$ are much lower. Also, Kovtyukh et al. overestimate
$\log g$ (relative to our values) by 0.5--0.7 dex for the two
F8 supergiants HR 2693 ($\delta$ CMa) and HR 7796 ($\gamma$ Cyg). 

In additon to this comparison with Kovtyukh et al.'s results, we
compared our data with about a dozen works. Setting aside the
details, we repeat the assertion made in the Introduction: the
accuracy of previous estimates of the fundamental parameters
may be inadequate for a reliable abundance analysis of
A-, F-, and G-type supergiants. Our determinations of the
parameters, and, in particular, the surface gravities determined from the
stellar parallaxes, will help to improve knowledge of the chemical
compositions of these stars.

Recently van Belle et al. (2009) published effective temperatures of cool supergiants
determined from near-infrared interferometry. 
These $T_{\rm eff}$ values are based on angular diameter measurements and published data on the 
narrow- and wide-band photometry in the 0.3 to 30 $\mu$m range, as well as the known spectral 
subtypes and luminosity classes of the stars. Moreover, distances of the stars are needed 
to compare the observed energy distribution with the model one; they were obtained from 
stellar parallaxes.
Four Ib stars are common to our investigations: $\alpha$ Per (HD~20902),
$\beta$ Cam (HD~31910), $\gamma$ Cyg (HD~194093) and
9 Peg (HD 206859). Van Belle et al.'s $T_{\rm eff}$s are
systematically higher ranging from 354 K for the hottest star $\alpha$
Per to 162 K for the coolest star 9 Peg. In the example of $\alpha$ Per 
presented in Table 1, the eight cited papers report effective
temperatures ranging from 6200 K to 6541 K, all lower than
van Belle et al.'s estimate of 6704$\pm$36 K. Examples of larger
overestimates may be cited, e.g., van Belle et al.'s $T_{\rm eff}$s
for the G0II stars $\epsilon$~Leo (HD~84441) and $\alpha$~Sge (HD~185758) are
6645$\pm$40~K and 6581$\pm$59~K, respectively. Not only are these
values more than 1000 K hotter than our estimates for this spectral
type (see Figure 3) but at least five other publications report
$T_{\rm eff}$ values for this pair of supergiants in the interval
5300--5400 K in good agreement with our scale but lower by
1200--1300 K than van Belle et al.'s values. 
In particular, the effective temperature of $\alpha$~Sge has been found from the IRFM, 
namely $T_{\rm eff} = 5415\pm 38$ K (Blackwell \& Lynas-Gray, 1998). 

Therefore, van Belle et al.'s effective temperatures $T_{\rm eff}$ for supergiants tend to be 
overestimated, and the overestimation can be rather great for some stars like $\epsilon$~Leo and $\alpha$~Sge. 
In order to understand a cause of this discrepancy, it is necessary to analyse in detail all 
observational data used by van Belle et al. In particular, we compared their distances $d$ with 
values inferred from new parallaxes of van Leeuwen (2007) and concluded that there are no 
significant differences (e.g., there is a good agreement for $\epsilon$~Leo and $\alpha$~Sge). Next we 
compared their interstellar extinction $A_V$ with our $A_V$ values determined from $uvby$ photometry 
and found that their $A_V$ are systematically overestimated. As a result, their reddening-corrected 
energy distributions correspond to $T_{\rm eff}$'s that are higher then ours. A complete analysis of 
possible errors in van Belle et al's method is outside the present work.

\section{Stellar mass and age}

\begin{figure}
\epsfxsize=8truecm
\epsffile{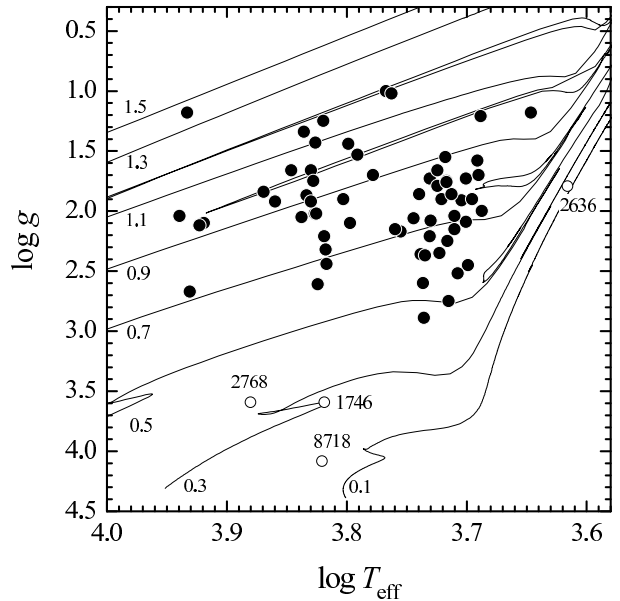}
\caption{The $T_{\rm eff}$ -- $\log g$ diagram. Evolutionary tracks
from Claret (2004) are shown and labelled by $\log M/M_\odot$. Programme
stars are plotted with filled circles. The four stars from Table 5 are
labelled next to unfilled circles. }
\end{figure}

Stellar mass and age are determined from the ($T_{\rm eff},\log g$)
and the evolutionary tracks of Claret (2004). In Figure 5, the
supergiants are represented by filled circles. The four stars shown by
unfilled circles are those shown by us not to be FG supergiants
(see Table 5); Figure 5 displays in striking fashion that they
are far outside our region of interest. In particular, masses of these four
stars are too low (see as well the last column of Table 5).

A characteristic of the evolutionary tracks for masses $M > 6M_\odot$
is the occurrence of extended loops returning stars from
the red supergiant branch. (For $M < 6M_\odot$, the loops are
much less pronounced.) Thanks to the loops, the stars appearing
as A, F and G supergiants may be in one of the three passes 
into the Hertzsprung gap. This leads to an ambiguity
in the assignment of a mass to a particular star. Fortunately,
the second and third crossings of the gap are almost
superimposed. The separation between the first and
second/third crossings corresponds to an ambiguity in mass of
about 25\%, i.e., a star actually on the second/third crossing but
considered to on the first crossing is given a mass that is
about 25\% too high. Lyubimkov et al. (2009) considered this
matter in detail for $\alpha$ Lep  (F0Ib) for which we assign the
mass of $14M_\odot$. The difference in mass and gravity between
the crossings corresponds to 0.07 dex or 18\%. In deriving the
mass $M$ (and the surface gravity $\log g$), we assume that the
supergiants are on their first crossing of the Hertzsprung gap.

A change of surface composition is predicted for cool
supergiants toward the end of the first crossing  and
again near the beginning of the third crossing. These
changes driven by a deep convective envelope are
the so-called first and second dredge-ups. 
These dredge-ups mix CNO-cycled products to the
atmosphere. Our abundance analyses to be reported in
later papers will provide a way to discriminate between
stars on first, second and possibly the third crossings of
the gap. 
 Rotationally-induced mixing in the main sequence
phase may also affect the surface compositions and thus
complicate the discrimination between the three crossings. 

The derived masses $M/M_\odot$ and ages $t$ are presented in Table 7. The
errors in  $M/M_\odot$ are estimated from uncertainties in parameters $T_{\rm eff}$
and $\log g$. One may see from Table 7 that 
a majority (57 of 63) of the supergiants have masses between 4$M_\odot$ and
15$M_\odot$ corresponding to early to middle B-type stars on the
main sequence. The most massive star, HR 825 with $M = 24M_\odot$,
began its evolution as a O-type main sequence star. Five stars
have masses near $3M_\odot$ and were late B-type main sequence stars.
All stars are young.

$T_{\rm eff}$ and $\log g$.
\begin{table}
 \centering
 \begin{minipage}{80mm}
  \caption{Microturbulent parameter $V_{t,}$, iron abundance $\log\epsilon$(Fe), mass $M$ 
and age $t$}
  \begin{tabular}{cccccc}
 \hline
 HR &  HD & $V_{t,}$    & $\log\epsilon$(Fe) & $M/M_{\odot}$ & $t$,\\
    &     & km s$^{-1}$ &                    &               & 10$^{6}$ yr\\   
\hline
27   & 571    & 3.6$^{a)}$ & 7.41 $\pm$ 0.05 & 6.1 $\pm$ 0.4 & 62\\
157  & 3421   & 2.5        & 7.41 $\pm$ 0.12 & 4.8 $\pm$ 0.2 & 109\\
292  & 6130   & 2.7        & 7.55 $\pm$ 0.10 & 7.1 $\pm$ 1.4 & 44\\
461  & 9900   & 2.8:       & 7.47 $\pm$ 0.19 & 9.5 $\pm$ 1.2 & 25\\
792  & 16780  & 2.9$^{a)}$ & 7.49 $\pm$ 0.14 & 5.0 $\pm$ 0.6 & 100\\
849  & 17818  & 2.2:       & 7.58 $\pm$ 0.13 & 6.5 $\pm$ 0.9 & 54\\
1017 & 20902  & 5.3        & 7.43 $\pm$ 0.09 & 7.3 $\pm$ 0.3 & 41\\
1135 & 23230  & 3.5:       & 7.56 $\pm$ 0.06 & 4.8 $\pm$ 0.1 & 109\\
1242 & 25291  & 3.2        & 7.43 $\pm$ 0.11 & 8.3 $\pm$ 0.9 & 32\\
1270 & 25877  & 1.7:       & 7.59 $\pm$ 0.13 & 5.7 $\pm$ 0.5 & 73\\
1303 & 26630  & 3.6        & 7.41 $\pm$ 0.12 & 7.0 $\pm$ 0.4 & 46\\
1327 & 27022  & 1.2        & 7.41 $\pm$ 0.10 & 2.8 $\pm$ 0.2 & 434\\
1603 & 31910  & 4.8        & 7.46 $\pm$ 0.10 & 6.5 $\pm$ 0.4 & 53\\
1740 & 34578  & 4.3        & 7.42 $\pm$ 0.12 & 8.8 $\pm$ 2.2 & 29\\
1829 & 36079  & 1.3        & 7.41 $\pm$ 0.07 & 3.5 $\pm$ 0.1 & 240\\
1865 & 36673  & 3.9        & 7.53 $\pm$ 0.08 &13.9 $\pm$ 0.8 & 13\\
2000 & 38713  & 2.1$^{a)}$ & 7.40 $\pm$ 0.19 & 3.9 $\pm$ 0.5 & 184\\
2453 & 47731  & 2.3:       & 7.62 $\pm$ 0.16 & 6.5 $\pm$ 1.1 & 54\\
2693 & 54605  & 7.0        & 7.51 $\pm$ 0.09 &14.9 $\pm$ 1.6 & 12\\
2786 & 57146  & 3.2        & 7.55 $\pm$ 0.09 & 5.9 $\pm$ 0.5 & 67\\
2833 & 58526  & 4.0        & 7.58 $\pm$ 0.13 & 4.7 $\pm$ 0.5 & 113\\
2881 & 59890  & 5.2        & 7.33 $\pm$ 0.10 & 7.3 $\pm$ 0.5 & 42\\
3045 & 63700  & 5.1$^{a)}$ & 7.43 $\pm$ 0.15 & 9.9 $\pm$ 1.0 & 23\\
3073 & 64238  & 3.5        & 7.60 $\pm$ 0.06 & 4.2 $\pm$ 0.3 & 151\\
3102 & 65228  & 3.7        & 7.61 $\pm$ 0.07 & 5.1 $\pm$ 0.4 & 92\\
3183 & 67456  & 3.5        & 7.54 $\pm$ 0.12 & 5.4 $\pm$ 0.5 & 83\\
3188 & 67594  & 3.3        & 7.51 $\pm$ 0.12 & 6.6 $\pm$ 0.4 & 52\\
3229 & 68752  & 2.3        & 7.51 $\pm$ 0.08 & 5.2 $\pm$ 0.2 & 90\\
3459 & 74395  & 3.5        & 7.53 $\pm$ 0.13 & 5.2 $\pm$ 0.3 & 89\\
4166 & 92125  & 2.7        & 7.52 $\pm$ 0.11 & 4.2 $\pm$ 0.1 & 148\\
4786 & 109379 & 1.5        & 7.60 $\pm$ 0.07 & 3.7 $\pm$ 0.1 & 206\\
5143 & 119035 & 1.0        & 7.32 $\pm$ 0.13 & 3.2 $\pm$ 0.2 & 316\\
5165 & 119605 & 2.5        & 7.33 $\pm$ 0.11 & 4.2 $\pm$ 0.2 &152\\
6081 & 147084 & 2.8        & 7.53 $\pm$ 0.17 & 8.7 $\pm$ 1.3 & 29\\
6536 & 159181 & 3.0        & 7.52 $\pm$ 0.10 & 6.0 $\pm$ 0.2 & 65\\
6978 & 171635 & 4.6        & 7.41 $\pm$ 0.08 & 8.2 $\pm$ 0.5 & 33\\
7164 & 176123 & 2.5$^{a)}$ & 7.40 $\pm$ 0.15 & 4.5 $\pm$ 0.5 & 128\\
7264 & 178524 & 3.2        & 7.33 $\pm$ 0.09 & 5.9 $\pm$ 0.3 & 67\\
7456 & 185018 & 2.8        & 7.34 $\pm$ 0.12 & 5.5 $\pm$ 0.5 & 79\\
7542 & 187203 & 4.2:       & 7.67 $\pm$ 0.08 & 5.3 $\pm$ 0.7 & 86\\
7795 & 194069 & 3.1$^{a)}$ & 7.5  $\pm$ 0.14 & 5.3 $\pm$ 0.3 & 87\\
7796 & 194093 & 5.2        & 7.46 $\pm$ 0.06 &14.5 $\pm$ 1.1 & 12\\
7834 & 195295 & 3.6        & 7.50 $\pm$ 0.07 & 5.3 $\pm$ 0.4 & 85\\
8232 & 204867 & 3.7        & 7.60 $\pm$ 0.12 & 6.4 $\pm$ 0.3 & 56\\
8313 & 206859 & 2.8:       & 7.48 $\pm$ 0.14 & 7.1 $\pm$ 0.4 & 43\\
8412 & 209693 & 2.3$^{a)}$ & 7.55 $\pm$ 0.13 & 4.2 $\pm$ 0.3 & 152\\
8414 & 209750 & 3.8        & 7.53 $\pm$ 0.09 & 6.5 $\pm$ 0.3 & 53\\
8692 & 216206 & 3.4$^{a)}$ & 7.40 $\pm$ 0.13 & 5.6 $\pm$ 0.4 & 74\\
\hline
207  & 4362   & 4.0  & 7.38 $\pm$ 0.18 & 7.9  $\pm$ 2.2 & 36\\
825  & 17378  & 10.8 & 7.43 $\pm$ 0.09 & 23.9 $\pm$ 5.8 & 7\\
2597 & 51330  & 3.3: & 7.32 $\pm$ 0.13 & 7.1  $\pm$ 2.2 & 44\\
2839 & 58585  & 2.0: & 7.40 $\pm$ 0.14 & 8.6  $\pm$ 2.1 & 30\\
2874 & 59612  & 7.8  & 7.52 $\pm$ 0.18 & 12.9 $\pm$ 2.7 & 15\\
2933 & 61227  & 2.7  & 7.37 $\pm$ 0.13 & 7.0  $\pm$ 2.1 & 45\\
3291 & 70761  & 3.9: & 7.41 $\pm$ 0.16 & 14.2 $\pm$ 3.5 & 13\\
6144 & 148743 & 4.8: & 7.39 $\pm$ 0.16 & 10.0 $\pm$ 3.6 & 23\\
7014 & 17294  & 4.6  & 7.43 $\pm$ 0.15 & 10.0 $\pm$ 3.2 & 22\\
7094 & 174464 & 3.4  & 7.32 $\pm$ 0.16 & 9.1  $\pm$ 1.9 & 26\\
7387 & 182835 & 4.4  & 7.47 $\pm$ 0.11 & 12.5 $\pm$ 1.8 & 15\\
7770 & 193370 & 5.0  & 7.28 $\pm$ 0.07 & 10.0 $\pm$ 1.3 & 22\\
\end{tabular}
\end{minipage}
\end{table}
\begin{table}
\setcounter{table}{7}
\addtocounter{table}{-1}
 \centering
 \begin{minipage}{80mm}
  \caption{ -- continued}
  \begin{tabular}{cccccc}
 \hline
 HR &  HD & $V_{t,}$    & $\log\epsilon$(Fe) & $M/M_{\odot}$ & $t$,\\
    &     & km s$^{-1}$ &                    &               & 10$^{6}$ yr\\   
\hline
7823 & 194951 & 4.2  & 7.37 $\pm$ 0.12 & 7.8  $\pm$ 2.0 & 36\\
7847 & 195593 & 4.1  & 7.44 $\pm$ 0.15 & 11.2 $\pm$ 2.7 & 18\\
7876 & 196379 & 3.4  & 7.29 $\pm$ 0.16 & 10.6 $\pm$ 1.1 & 20\\
\hline
\end{tabular}
$^a)$ These $V_t$ values are  evaluated from mean relations in Figure 7. \\
\end{minipage}
\end{table}

\section{The microturbulent parameter}

The microturbulent parameter $V_t$ is determined by a standard
method. The accepted value is that which provides
abundance estimates from lines of some atom or ion that are
not dependent on the equivalent width $W$ of the lines. 
Model atmospheres used in the exercise are computed using
Kurucz's (1993) code ATLAS9 with our derived parameters

For this purpose, we choose the Fe\,{\sc ii} lines rather
than the Fe\,{\sc i} lines. Both atom and ion provide an
adequate number of lines. However, appreciable departures from
local thermodynamic equilibrium (LTE) may occur for Fe\,{\sc i}
lines. Such departures were first shown for F supergiants by
Boyarchuk et al. (1985) and confirmed  later for F and G stars
by other authors (e.g., Th\'{e}venin \& Idiart, 1999). These
departures compromise the use of Fe\,{\sc i} lines as a $V_t$
measure. Similar departures are anticipated for neutral atoms
of other iron-group elements (Lyubimkov et al. 2009). 
Lines of the ions are expected to be insensitive to departures
from LTE.
Schiller \& Przybilla (2008) confirmed with quantitative calculations
for the A2 supergiant $\alpha$ Cyg (Deneb) that the Fe\,{\sc ii} lines
are  not seriously affected by departures from LTE.  
Indeed, they showed by using Fe\,{\sc ii} lines with
$W$ up to 400 m\AA\  that $V_t$ is independent of depth in the
atmosphere. Lyubimkov \& Samedov (1990) had earlier suggested that
$V_t$ did vary with depth in the case of F supergiants but this
conclusion was based on Fe\,{\sc i} lines and requires
verification using Fe\,{\sc ii} lines.

\begin{figure}
\epsfxsize=8truecm
\epsffile{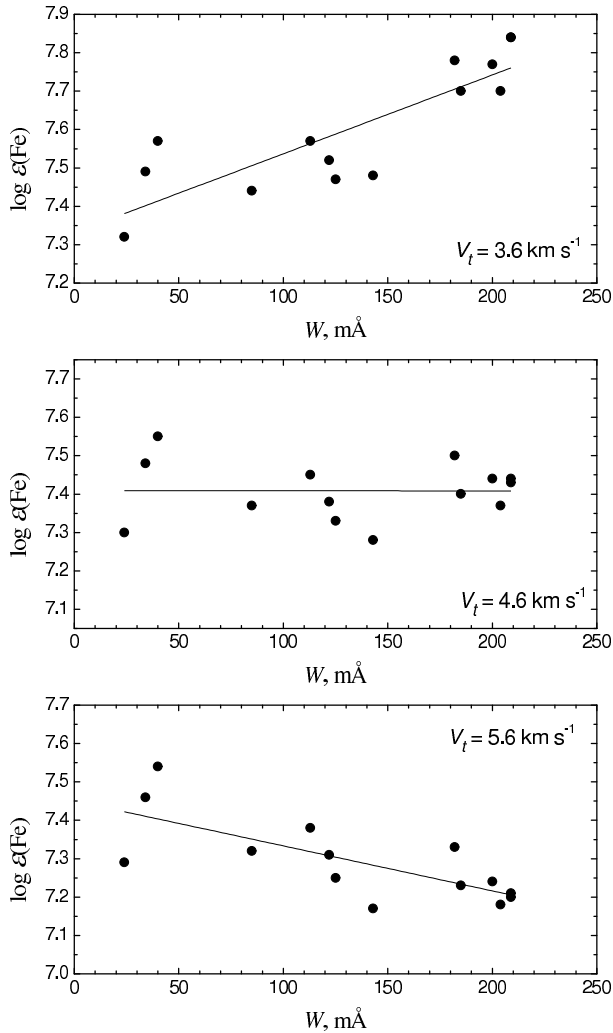}
\caption{The determination of $V_t$ for the star 45 Dra (HR 6978). The
iron abundance $\log\epsilon$(Fe) is shown as a function of the
observed equivalent width $W$ of Fe\,{\sc ii} lines for three different
$V_t$ values; $V_t$ = 4.6 km s$^{-1}$ is the adopted value.}
\end{figure}

\begin{figure}
\epsfxsize=8truecm
\epsffile{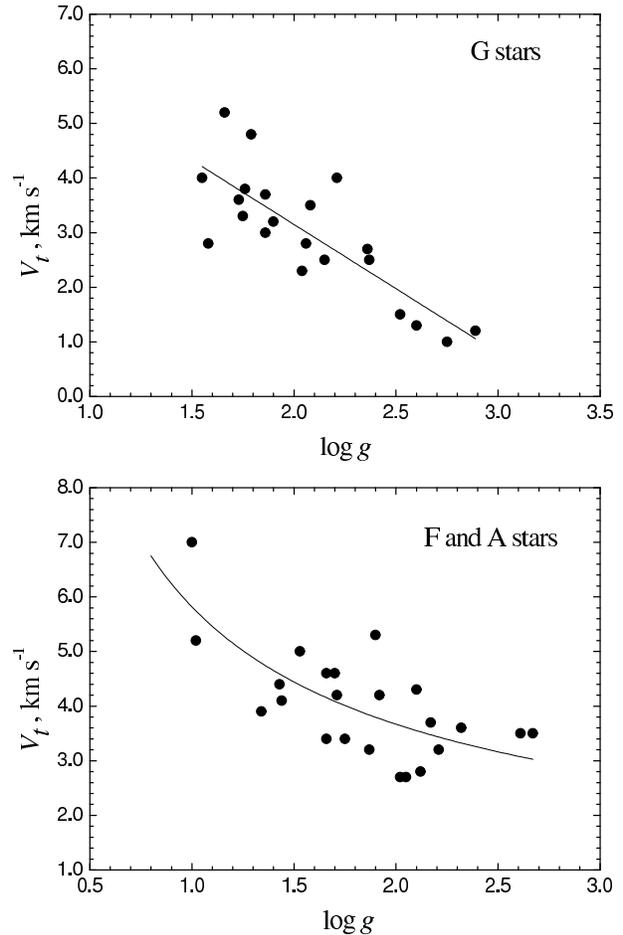}
\caption{Relations between the microturbulent parameter $V_t$ and
surface gravity $\log g$ for G supergiants (top panel) and A and F
supergiants (bottom panel). The solid lines are mean relations that
are used to estimate $V_t$ for those stars for which $V_t$ is
not reliably determined from Fe\,{\sc ii} lines.}
\end{figure}

The microturbulent parameter $V_t$ determination from Fe\,{\sc ii}
lines uses atomic data including excitation potentials and
oscillator strengths taken from the VALD database (Kupka et al.
1999; Heiter et al. 2008). For most stars, Fe\,{\sc ii} lines with $W < 250$ m\AA\
were used but, in order to obtain an adequate sample of lines, the
upper limit was $W = 450$m\AA\ for some A-type supergiants. 
Figure 6 shows abundance versus $W$ plots for  Fe\,{\sc ii} lines
from the F-type supergiant 45 Dra (HR 6978). The middle panel
is for $V_t = 4.6$ km s$^{-1}$, the adopted value.   The upper and
lower panels show the results of changing $V_t$ by $\pm$1 km s$^{-1}$.
We estimate that the uncertainty in $V_t$ is $\pm$0.5 km s$^{-1}$. The iron
abundance here and subsequently is given in the standard logarithmic
scale where the hydrogen abundance is $\log\epsilon$(H) = 12.00.

Derived $V_t$ values are presented in Table 7 with near stars ($d < 700$ pc)
and distant ones ($d > 700$ pc) displayed separately. Three categories of
$V_t$ estimates are distinguished in the table: Values determined
from Fe\,{\sc ii} lines by the method illustrated by Figure 6 to an
accuracy of about $\pm0.5$ km s$^{-1}$; values determined by this
method but less reliably so (e.g., some G-type supergiants have
few weak lines) and marked by a colon in the table; and  values
estimated from mean relations between $V_t$ and $\log g$ (see below)
and marked by the superscript `a'.

Our most reliable $V_t$ values are shown in Figure 7 as a
function of surface gravity for  G supergiants  (upper panel)
and A and F supergiants (lower panel). One may see that there 
is an obvious trend in both cases: $V_t$ tends to increase with 
decreasing $\log g$. For the G supergiants we fit a straight line 
by the least-squares method, but a more complex curve seems appropriate 
for the A and F supergiants. These mean relations were used to estimate 
$V_t$ when the standard method could not be applied because of a paucity 
of weak unblended lines. The scatter about these relations is somewhat 
larger than the estimated error of $\pm$0.5 km s$^{-1}$ of a $V_t$ determination.
Such empirical relations cannot be strictly valid. In fact, the
relations do not express fully the likely temperature
dependence. In particular, the two hottest A supergiants,
HR 825 and HR 2874, are clearly outliers; their $V_t$ values of 10.8 and 
7.8 km s$^{-1}$, respectively, would be offscale in the lower panel of Figure 7. 

\begin{figure}
\epsfxsize=8truecm
\epsffile{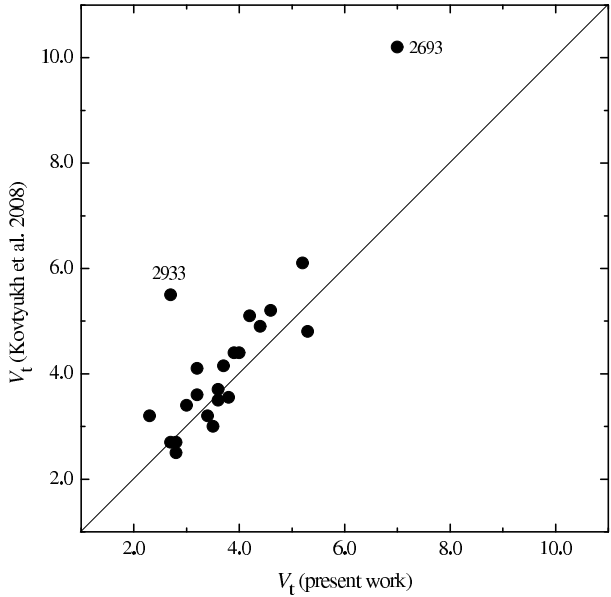}
\caption{Comparison of our $V_t$ values with those of Kovtyukh et al.
(2008). The solid straight line denotes perfect agreement between the
two datasets.}
\end{figure}

A comparison with Kovtyukh et al.'s (2008) microturbulent parameter,
also derived from Fe\,{\sc ii} lines, is possible for
the 22 common stars in Figure 8. For 20 of the 22 stars, differences between our
and their estimates are within $\pm1.0$ km s$^{-1}$ and the mean
difference is only 0.4 km s$^{-1}$. For two F supergiants, HR 2693 
and HR 2933, the difference is about 3 km s$^{-1}$. It may be significant
that the Kovtyukh et al,'s values for the parameters $T_{\rm eff}$ and $\log g$ 
differ appreciably from ours for these two stars.

\section{Iron abundance and metallicity}

The iron abundance is necessarily obtained  the same time as the  
microturbulent parameter is found - see Figure 6. The
abundance $\log\epsilon$(Fe) for each star  is
given in Table 7. The mean error in the abundance also given
there incorporate  the scatter in the abundance estimates from
the various Fe\,{\sc ii} lines and the uncertainties in the
parameters $T_{\rm eff}$ and $\log g$.   
The average error for the local ($d < 700$ pc) supergiants is
$\pm$0.11 dex and $\pm$0.14 dex for the distant ($d > 700$ pc)
supergiants. 

\begin{figure}
\epsfxsize=8truecm
\epsffile{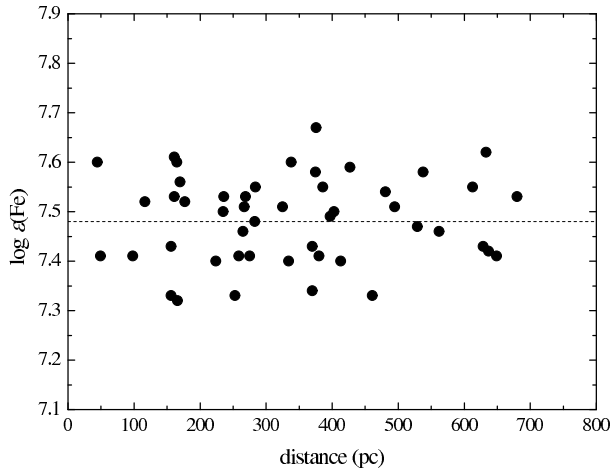}
\caption{Iron abundance of the local ($d < 700$ pc) AFG supergiants
as a function of their distance $d$. The dashed line corresponds to the
mean abundance $\log\epsilon$(Fe) = 7.48.}
\end{figure}

The mean iron abundance for the 48 local supergiants is 
$\log\epsilon$(Fe) = $7.48\pm0.09$ and 7.39$\pm0.07$ for the
15 distant supergiants. The most distant stars, HR 825 and HR 3291 at $d=2700-2900$ pc, 
have the abundances $\log\epsilon$(Fe) = 7.43$\pm0.07$ and 7.41$\pm0.14$, respectively. 
The iron abundance is independent of distance: Figure 9 shows
the abundance versus distance plot for the local sample.
These iron abundances are equal to the solar abundance which is
$\log\epsilon$(Fe)= 7.45$\pm$0.05 (Grevesse et al. 2007). 

As mentioned above, when analyzing Fe\,{\sc ii} lines, we used atomic data
including oscillator strengths $gf$ from the database VALD. Recently a
new set of accurate $gf$-values for Fe\,{\sc ii} lines was published by
Mel\'endez \& Barbuy (2009). We compared two sets of $gf$-values for our 
Fe\,{\sc ii} line list and found that (i) there is very good agreement in $gf$ for
majority of individual lines and (ii) there is no notable systematic
difference between two sets. Therefore, all our conclusions concerning
the microturbulent parameter $V_t$ and the iron abundance $\log\epsilon$(Fe) remain
in force. 

\section{Concluding remarks}

In this paper, we have set the stage for a new abundance analysis of  
AFG supergiants of
luminosity classes Ib and II from the solar neighbourhood.  
The major contribution of
this paper is our determination  of accurate
fundamental parameters characterizing the stellar 
atmospheres: $T_{\rm eff}, \log g$, $V_t$, and $\log\epsilon$(Fe).  
Of  particular note is our use of  stellar parallaxes
from the rereduction of the {\it Hipparcos} parallaxes (van Leeuwen 2007) in 
determining the surface gravity $\log g$ with what is claimed to be an
unprecedented precision: the error is typically $\pm$0.06 dex 
for supergiants with
distances $d < 300$pc increasing to $\pm0.12$ dex for the supergiants with
$d$ between 300 pc and 700 pc. In the case of  supergiants with distances greater
that 700 pc, the parallaxes are too small or unmeasureable for the stellar
parallax to provide a useful constraint on the gravity and, then, the typical error
in $\log g$ is 0.2--0.3 dex, the commonly declared accuracy for published
$\log g$ values.  The primary $T_{\rm eff}$ indicators are the Balmer lines and 
the indices $[c_1]$, $Q$ and $\beta$ 
which provide loci in the plane $(T_{\rm eff},\log g$) plane with the
degeneracy broken by the  $T_{\rm eff}$-insensitive  $\log g$ estimate from the
stellar parallax. The mean error in $T_{\rm eff}$ for stars with $d < 700 $ pc is
$\pm120$ K. Our accurate $T_{\rm eff}$ for  63 supergiants, some of which are
reclassified on the basis of our effective temperatures and gravities,  are the basis
for a new $T_{\rm eff}$ scale for A5--G5 stars of luminosity 
classes Ib--II.
(Three stars  selected from the {\it Bright Star Catalogue} as A or F supergiants are
 shown to be dwarfs or subgiants. A fourth star listed as a G supergiants is shown to
be a  K supergiant. This quartet  will not be considered in subsequent papers.)

The turbulent parameter $V_t$ and the iron abundance $\log\epsilon$(Fe)
were determined from Fe\,{\sc ii} lines on account of their insensitivity to
non-LTE effects. The parameter $V_t$ is seen to be gravity dependent:
$V_t$  increases with decreasing $\log g$.  The iron abundance is independent
of stellar distance $d$ and equal to the solar Fe abundance: the mean abundance
for the 48 supergiants with distance $d < 700$ pc is $\log\epsilon$(Fe) = 7.48$\pm$0.09
dex, a value coincident with the solar abundance $\log\epsilon$(Fe) = 7.45$\pm$0.05
dex (Grevesse et al. 2007).

The coincidence between the iron abundance of young nearby stars and that of the 4.5~Gyr 
old Sun is interesting from the viewpoint of models of Galactic Chemical Evolution 
(GCE). It is important to note that there are other data as well, which confirm a closeness of 
the metallicity of these stars to the solar one. We may mention our earlier conclusion based on 
the Mg abundance in local B stars, the progenitors of the AFG supergiants. 
We reported the abundance $\log\epsilon$(Mg) = 7.59$\pm$0.15 (Lyubimkov et al. 2005), a value 
equal within the uncertainties to the solar value of $\log\epsilon$(Mg) = 7.53$\pm$0.09.  Moreover,
studies of young stars, both hot and cool,  by a variety of authors have shown that their
compositions are very similar to that of the Sun (see, e.g., the iron abundance obtained 
by Luck, Kovtyukh \& Andrievsky (2006) for Cepheids in the solar neighbourhood, the iron and magnesium 
abundances found by Fuhrmann (2004) for nearby F, G and K dwarfs and subgiants of the thin Galactic 
disk, and the sulfur abundance derived dy Daflon et al. (2009) for B-type main-sequence stars in the 
Orion association).

The question arises: may these results be reconciled with models of GCE? One may cite the 
recent work of Spitoni et al. (2009), where an enrichment of the solar neighborhood by various 
metals is studied, in particular, by Fe and Mg (see their Fig. 16 and 17). One sees from these 
results that during the Sun's life the Fe and Mg abundances in its neighborhood are predicted to 
increase by about 0.15 dex and 0.05-0.07 dex, respectively. The ordinary accuracy of observed 
abundances in stars seems to be insufficient to detect such a small enrichment. In particular, 
according to our above-mentioned finding, the enrichment 
$[Fe/H]$ = 0.03$\pm$0.09 and $[Mg/H]$ = 0.06$\pm$0.15 takes place, that does not contradict 
Spitoni et al.'s theoretical estimations but is negligible in comparison with uncertainties in the 
derived Fe and Mg abundances.

The fundamental parameters  
and Claret's (2004) evolutionary tracks provide 
estimates of stellar mass and age.  The great majority, 57 of the 63 stars, have
masses between 4$M_\odot$ and 15$M_\odot$  showing that their progenitors
were early to middle B-type main sequence stars. Five of the remaining six stars
have lower mass, $M =2.8-3.9M_\odot$, and their progenitors were late B-type
main sequence stars.  The final star in the minority of six is HR 825 with an 
estimated mass of 24$M_\odot$ and its progenitor was an O-type main sequence
star. 

In subsequent papers, we shall derive detailed chemical compositions  for the
supergiants.  A primary aim of this future work, as noted in the Introduction, is to
quantify accurately the various signatures of internal mixing that are
anticipated to occur in preceding stages of evolution  from the main sequence to
the dredge-ups occurring in supergiants.  In particular, we shall compare the
compositions of the supergiants with the results for early and
middle B-type stars from our continuing collaboration (Lyubimkov et al.
2000, 2002, 2004, 2005).

\section{Acknowledgements}

We thank Dr. A. Korn for useful discussion. This research has been supported by  
the CRDF grant UKP1-2809-CR-06 and by  the grant F-634 from the Robert A. Welch 
Foundation of Houston, Texas.

\label{lastpage}

\end{document}